# Experimental investigation of twin pulsed jets in a hemispheric elastic cavity

L. S. Merlo[1], L. Kadem[1], W. Saleh[1], H. D. Ng[1], G. Di Labbio[2*]

[1] Department of Mechanical, Industrial and Aerospace Engineering, Concordia University, Montréal, QC, Canada

[2] Département de génie mécanique, École de technologie supérieure, Montréal, QC, Canada

[*]Corresponding author: giuseppe.dilabbio@etsmtl.ca

## ABSTRACT

This study experimentally examines the impact of spacing between two pulsed jets and their strengths on the fluid dynamics within an elastic hemispherical cavity. Such interactions between multiple pulsed jets are observed in various natural and industrial contexts, including cardiovascular flows, where they occur naturally within the atria or result from medical interventions (e.g., mitral valve repair, mechanical heart valves, paravalvular leaks) or diseases (e.g., aortic or pulmonary valve regurgitation). Fundamentally, these flows usually feature two or more pulsed jets interacting in an expanding, elastic environment. In this investigation, the experimental setup features two parallel pulsed jets entering the cavity, with jet strength varied across five formation times (1, 2, 3, 4, 5) and four spacing ratios (1.5, 2.0, 2.5, 3.0). Time-resolved particle image velocimetry is used to capture the instantaneous velocity fields. The results reveal three distinct flow regimes: short-time decay, decay at the wall, and wall rebound—with or without the formation of secondary vortices. These findings uncover rare aspects of twin vortex ring behavior, including symmetry breaking, trajectory shifts, and wall-induced rebound mechanisms, with direct relevance to cardiac fluid dynamics in both healthy and pathological conditions.

## I. INTRODUCTION

A pulsed jet refers to a burst of fluid ejected through a nozzle, which can generate coherent vortex ring structures when propagating into a relatively still environment. Pulsed jets are highly efficient in energy transfer, making them particularly attractive for applications ranging from underwater propulsion and maneuvering to the filling of heart cavities [1]–[6]. Despite the potential application of multiple pulsed jets in propulsion and locomotion, studies on the fluid dynamics resulting from multiple pulsed jet interactions in unconfined and confined spaces are scarce. In particular, little is known about the fluid dynamics of twin pulsed jets within a confined or walled space, with varying strengths and spacings, which motivates the present study.

Relevant studies to the present work include those by Athanassiadis and Hart [7] and Chevalier [8]. Athanassiadis and Hart [7] experimentally investigated the influence of multi-jet interactions on propulsion by examining the wakes generated by two parallel pulsed jets, in an unconfined space, using dye visualizations. They examined the mutual influence of these jets under low-speed maneuvering conditions (Reynolds number Re ≈ 350 and stroke ratio $L/d$ = 2) on thrust production and propulsive efficiency for different nozzle spacings, ranging from noninteractive jets to interactive jets. Their findings indicate that interactive jets in close proximity can have an unfavorable effect on propulsive efficiency by decreasing the thrust by up to 10% [7]. However, they have shown that for interjet spacings larger than 2.5 tube diameters, the effects of jet interaction on thrust become negligible. Chevalier [8] studied, in a similar configuration, the flow structures generated by parallel twin pulsed jets for different nozzle spacings ($s/d$) and stroke ratios ($L/d$). The selected Reynolds number (Re ≈ 3500) is more representative of the scale for aquatic vehicles and other multiple pulsed jet applications, such as cardiovascular flows. The study identifies a critical spacing ratio of 3 above which no vortex ring interactions are observed. When the spacing ratio falls below 1.5, the vortex rings merge at the nozzle exit. The nozzle spacing significantly influences the reconnection and merging point of vortex rings, particularly for $s/d < 2$. Yet important to understand the flow dynamics of multiple pulsed jets, the aforementioned studies were performed in an unconfined space.

In the absence of systematic studies on the interaction of multiple pulsed jets with a confined, flexible boundary, one can only draw some conclusions based on equivalent scenarios with single pulsed jets. For instance, there has been an increasing interest in fundamental studies regarding the flow dynamics when a single jet impinges on complex static shapes, such as V-walls and hemispheres. Ahmed and Erath [9] investigated the dynamics of vortex rings impinging on a concave hemispherical cavity for vortex-to-hemisphere radius ratios between 1/4 and 2/3. Their experiment involved releasing a fluid mass (Re = 1450) from a single jet, with a circular orifice, onto a rigid hemispherical cavity. The setup allows the vortex ring to propagate around the cavity without being confined. As the radius ratio between the vortex ring and the cavity increases, there is a significant increase in vorticity transfer from the primary vortex ring to the secondary vortices. The largest ratio results in the secondary vortex circulation being approximately ten times greater than studies conducted on flat wall impact. This interaction also results in the explicit formation and detachment of a coherent ring, which then advects 180° from the initial path of the primary vortex ring. In a related study, New et al. [10] investigated the interaction of vortex rings with V-walls at various angles (0°, 30°, 60°, 90°, and 120°) and Reynolds numbers (Re = 2000 and 4000). They found that V-walls induce complex three-dimensional vortical changes, especially those with smaller angles. Moreover, primary, secondary, and tertiary ring core sizes decrease drastically and occur more rapidly as the angle reduces. Secondary and tertiary vortex ring cores leapfrog past primary vortex ring cores along the surfaces of the V-wall after their entrainment in cases involving high Reynolds numbers and/or smaller V-wall angles. Zhang et al. [11] examined a single vortex ring impinging on a concave hemispherical shell across Reynolds numbers ranging from 750 to 7000. They observed the formation of straight-edged secondary vortex pairs propagating upward after wall collision, with greater vortex ring extension along the straight edge compared to the concave direction, a consequence of their unidirectional confinement (half-cylinder geometry) [11]. Other studies also included flexible walls. Samaee [12] performed a study on the decay of a single vortex ring in flexible-wall spheroidal confined domains, including semi-oblate, hemispherical, and semi-prolate cavities. The experimental setup involves a single jet injecting fluid into a confined cavity, creating a vortex ring under varying Reynolds numbers (Re = 1400, 1700, and 2600). The results show that axial confinement has minimal impact on the formation number of vortex rings but significantly influences the peak circulation and decay rates.

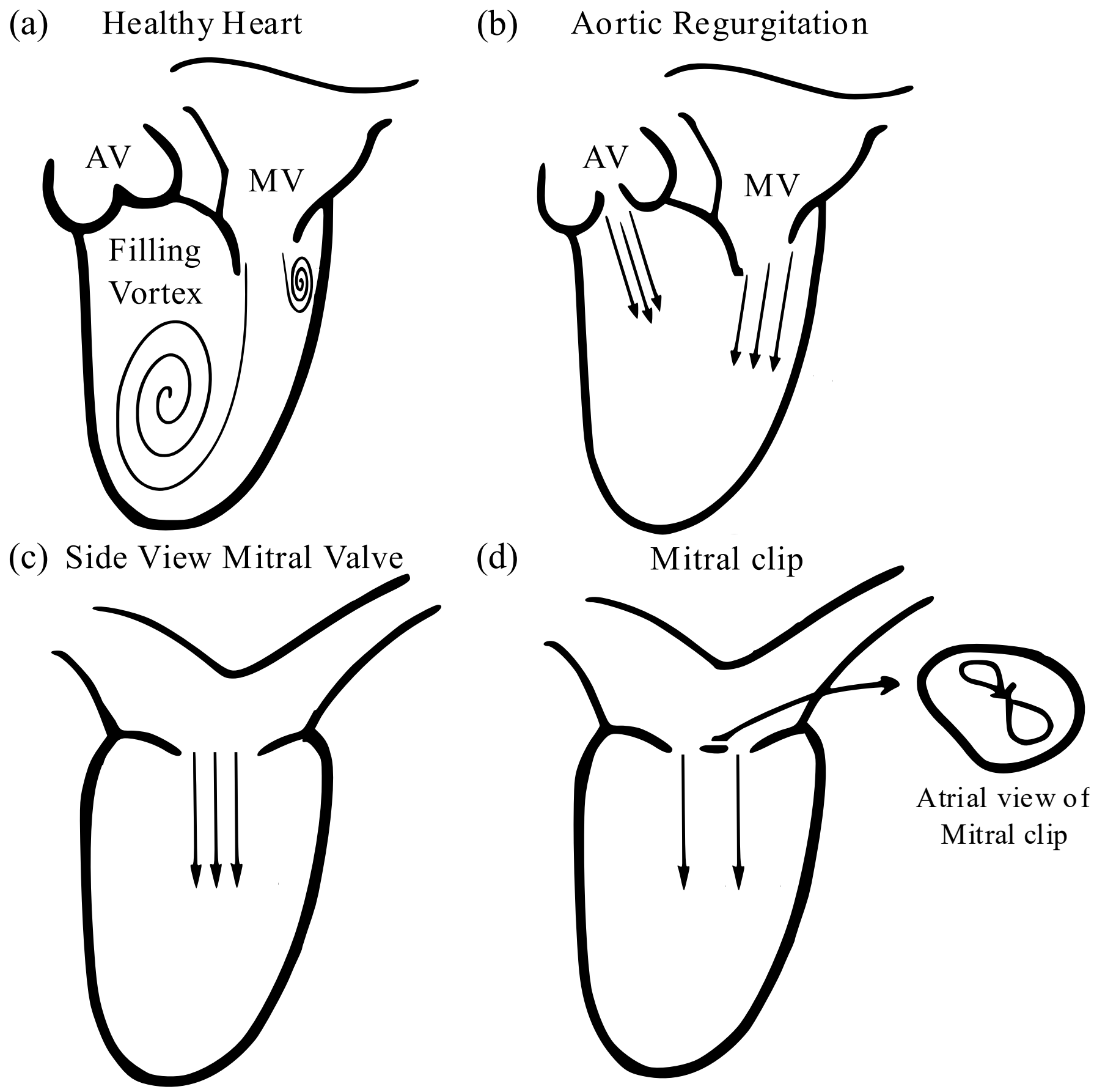

**FIG 1.** Schematic of some examples of flow configurations where twin pulsed jets appear in the cardiovascular system. (a) In a healthy left ventricle, inflow occurs solely through the mitral valve (MV) while the aortic valve (AV) remains closed. (b) In the presence of aortic valve regurgitation, the aortic valve (AV) leaks, leading to an abnormal filling of the left ventricle from both the mitral valve (MV) and the aortic valve (AV). (c) A side view of the left ventricle under healthy conditions, showing a single jet filling the ventricle through the mitral valve. (d) A side view of the left ventricle with a Mitral Clip or edge-to-edge repair of the mitral valve, resulting in two jets emerging into the left ventricle.

The fact that the existing literature on pulsed twin jets has mostly been limited to unconfined spaces or a few studies primarily on the vortex–wall interaction from single pulsed jets on static walls reveals a research gap in the fundamental understanding of the complex interaction of twin pulsed jets within a confined moving boundary domain. It is worth mentioning that such complex configurations are, in fact, very common in the context of cardiovascular fluid dynamics, where twin pulsed jets can be found in specific situations like edge-to-edge repair, aortic valve regurgitation [13], and pulmonary valve regurgitation [13] (See Fig. 1. For some illustrative examples). Edge-to-edge repair of mitral valve regurgitation leads to parallel twin pulsed jets in the left ventricle [14]. Teimouri et al. [14] experimentally investigated the flow in an elastic model of a left ventricle under three conditions: with a healthy mitral valve, with a regurgitant mitral valve, and following an edge-to-edge repair using a MitraClip device. This device alters the intraventricular flow dynamics by generating twin pulsed jets with a very small interjet distance. Their results showed that although the twin-jet configuration increases energy loss in the left ventricle compared to the healthy condition, it helps mitigate the adverse effects of mitral regurgitation. Aortic valve regurgitation also results in an interaction between two pulsed jets in the left ventricle. According to Di Labbio et al. [15], [16], this flow interaction leads to a sub-optimal blood flow that increases energy dissipation. Similar findings have been reported by Mikhail et al. [13] in the right ventricle in the context of pulmonary valve regurgitation. However, the jet interactions observed in these flows are indeed highly complex and difficult to interpret [15], [16].

To this end, the objective of the current experimental study is to advance fundamental understanding of the interaction between twin pulsed jets within a deforming elastic cavity. More specifically, the effects of both formation time and interjet spacing of twin parallel single-pulsed jets on the flow structures ejected within a hemispheric elastic cavity are explored. This paper is organized as follows. In Section II, the experimental setup and the phenomena observed via particle image velocimetry (PIV) are first described. Section III presents the experimental results and discussion. A more in-depth interpretation of the experimental flow field is provided from the analyses using the Lagrangian-averaged vorticity deviation (LAVD) method, showing three distinct, primary flow behaviors for all cases studied and contributing to a flow map illustrating the different flow configurations. Section IV continues with a discussion on the innovation and implications of the present work, summarizing some key findings of this study and offering some concluding remarks.

## II. METHODS

In this experimental study, the pulsed jets are produced via two nozzles with flat ends extending downward from a rigid wall in the horizontal plane. The nozzles consist of two parallel rigid tubes with an inner diameter ($d$) of 2.45 cm and an outer diameter of 3.05 cm within standard manufacturing tolerances. The tubes extend 27.23 cm vertically above the elastic transparent silicone hemispheric cavity, having an inner diameter ($D$) of 12.43 cm and a thickness of 2.15 $\pm$ 0.30 mm, enclosed within an acrylic tank. The cavity is molded by coating a polished 3D-printed hemisphere with silicone and curing it in a two-axis rotating heated curing chamber to ensure uniformity of the coating. The silicone has a shore hardness of 40A. The tubes are secured by holders that are 3D-printed using Polylactic Acid (PLA) material. The two tubes, the elastic cavity, and the rest of the tank are filled with a water-glycerol mixture with a volume ratio of 60% water and 40% glycerol. This working fluid was selected to match the refractive index of the silicone hemisphere (n = 1.39) and to operate at a lower Reynolds number. The mixture has a density ($\rho$) of 1100 kg/m$^3$ and a dynamic viscosity ($\mu$) of 4.2 cP.

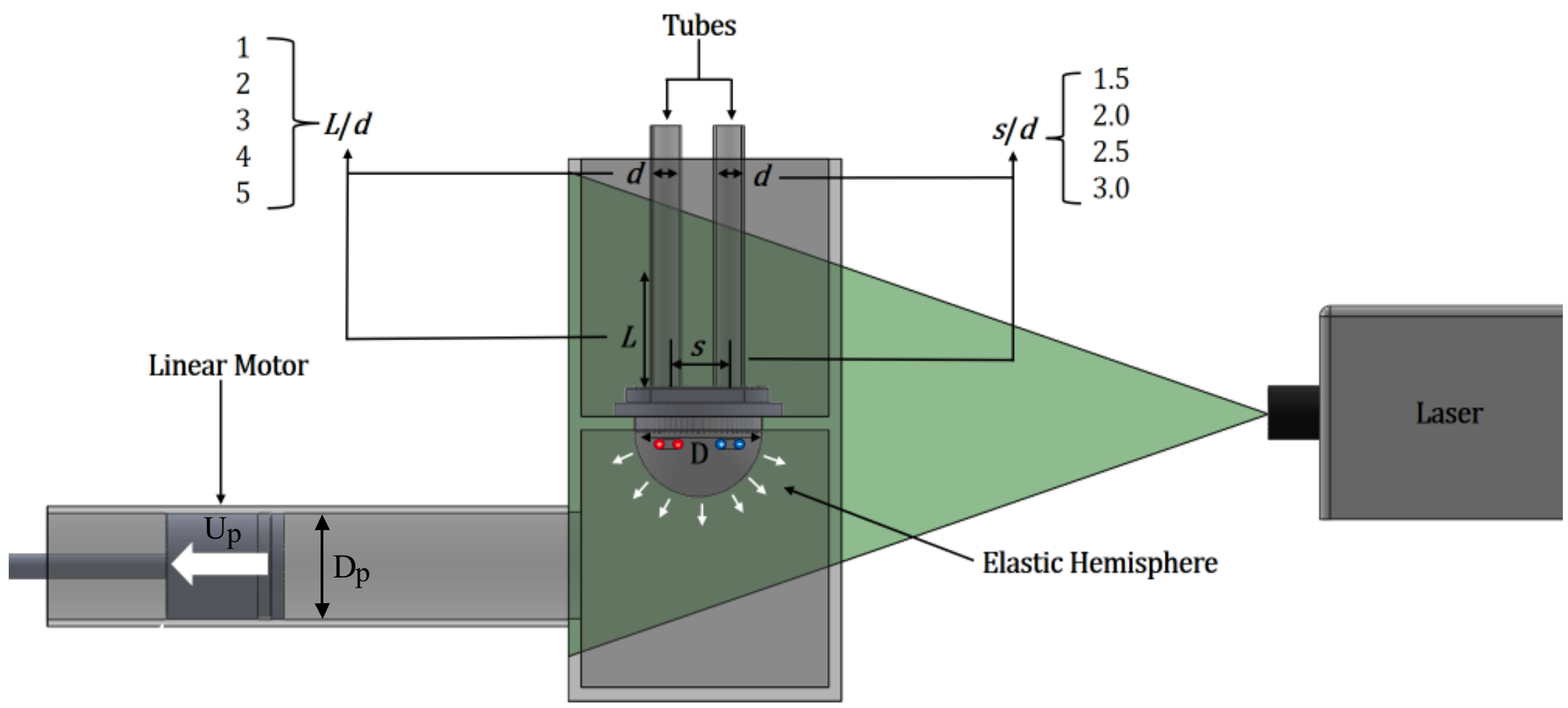

**FIG 2.** Schematic of the overall setup showing the retraction of the linear motor which forces the expansion of the hemispherical elastic cavity and, consequently, the ejection of the pulsed jets from the indicated tubes. Different formation times ($t_f^*$ = $L/d$ = 1, 2, 3, 4, and 5) and spacings ($s/d$ = 1.5, 2.0, 2.5, and 3.0) are investigated.

As seen in Fig. 2, on one end of the tank, a piston-cylinder is driven by a linear motor (LinMot PS01-37x120) using the LinMot Talk software (NTI AG; Switzerland). As the piston-cylinder moves backward at a specified velocity, it causes the elastic cavity to expand and, as a result, a pulsed jet emerges from each tube into the cavity. The height of the working fluid in the tubes was set to 10 cm above the rigid flat surface of the hemispheric cavity (and open to the atmosphere) at the start of each backward piston stroke. The height of the columns fell by 1.2 to 6.0 cm, depending on the piston stroke used in this study. The static jet exit pressure, therefore, varies by up to 648 Pa. The piston's velocity ($U_p$) on the system was set to vary between 1.2 and 6.0 cm/s, and the inner diameter of the cylinder ($D_p$) is 5 cm. The jet velocities ($U$) were computed using the prescribed piston velocity and the conservation of mass:

$$U = \frac{1}{2}\left(\frac{D_p}{d}\right)^2 U_p \tag{1}$$

The piston velocities were determined based on the desired formation times ($t_f^*$) for this study, which defines the ratio of the length of an ejected fluid column to its diameter [17]:

$$t_f^* = \frac{L}{d} = \frac{\overline{U}T}{d} \tag{2}$$

where

$$\overline{U} = \frac{L}{T} = \frac{1}{T}\int_0^T U(t)\mathrm{d}t \tag{3}$$

is the mean velocity discharged from the jets, $T$ the pulse time, $d$ the inner diameter of the tubes, and $L$ the length of the ejected fluid column. The different formation times ($t_f^*$) reproduced on the system are 1, 2, 3, 4, and 5. The piston is set to retract with $T$ = 1 s for all cases and sufficient time was implemented to allow all residual flow structures to be fully dissipated before the subsequent run. The piston stroke ($L_p = \overline{U}_p T$) to be determined from the formation time:

$$t_f^* = \frac{1}{2}\left(\frac{D_p}{d}\right)^2 \frac{L_p}{d} \tag{4}$$

Four different spacing ratios were chosen for this study, represented by $s/d$, where $s$ is the distance between the axes of the tubes and $d$ is the inner diameter of the tubes: $s/d$ = 1.5, 2.0, 2.5, and 3.0. Each case was recorded three times to examine repeatability. Using the peak vorticity of the emerging vortex rings as a reference, the differences between realizations were found to range from 3 to 10 %. The Reynolds number ($\mathrm{Re} = \rho\overline{U}d/\mu$) for the two parallel jets ranges from 75.72 to 385. For each condition, Particle Image Velocimetry (PIV) is used to capture the time-resolved 2D-2C velocity fields and to observe the flow interaction between the two pulsed jets within the hemispheric elastic cavity. The plane of interest is the symmetry plane of the elastic hemisphere bisecting the axes of the two nozzles. Within this plane, and given the low Reynolds numbers used in this study, the flow is expected to be two-dimensional. Polyamid seeding particles (Dantec Dynamics; Denmark) are used to visualize the flow (mean diameter: 50 µm, density: 1.03 g/cm$^3$). The particles are illuminated using a double-pulsed Nd:YLF laser (LDY301, Litron Lasers; England) having a wavelength of 527 nm. The images are captured using a high-speed dual-frame CCD camera (Phantom v310, Vision Research; USA) with a Nikon AF Micro-Nikkor 60 mm f/2.8D lens. The camera position and focus were adjusted to contain the whole hemispheric cavity in the field of view, while still ensuring particle image sizes of 2 to 3 pixels on average. The flow is recorded for one cycle when the piston is pulled back. The piston pulls back in 1 s, and the number of images taken for formation times of 1, 2, and 3 is 1200, while for those of 4 and 5, it is 1600 and 1200, respectively. The number of images for $t_f^*$ = 4 was increased because large coherent structures persisted in the flow field for a longer period, whereas they were rapidly dissipated at higher $t_f^*$. The velocity fields are computed in DaVis 8.4 (LaVision; Germany) using a multi-pass cross-correlation algorithm with decreasing window size in three passes from 64 × 64 px to 16 ×16 px, each using 50% overlap. The raw velocity vector fields were processed using DaVis 8.4 to eliminate spurious vectors, interpolate missing data, and smooth the flow. Any vectors with a correlation peak ratio below 1.2 were discarded. A 3 × 3 median filter with outlier detection was applied, removing vectors that exceeded 2 standard deviations from the local median. Additionally, isolated clusters containing fewer than 5 vectors were eliminated. Gaps in the data were interpolated, and a 3 × 3 spatial filter was applied twice to ensure continuity and reduce noise. The final spatial resolution is 0.88 × 0.88 mm. The double pulse time interval for the formation times of 1, 2, and 3 is 2400 µs, while for those of 4 and 5, it is 1200 µs and 1600 µs, respectively. The acquisition frequency for formation times 1, 2, and 3 is 200 Hz, while for those of 4 and 5, it is 400 Hz and 300 Hz, respectively. The total uncertainty in the velocity field is estimated to be below 5% with respect to the maximum pointwise velocities observed (0.62 m/s), based on the repeatability study and the major uncertainties described in [18], [19]. The mean velocity uncertainty, quantified using the correlation statistics method, is around 0.02 m/s, indicating a high level of measurement confidence across the field [20].

To aid in interpreting the flow field from experiments, the vortices were mapped through time using the Lagrangian-averaged vorticity deviation (LAVD) method. The LAVD is a local measure of material rotation relative to the spatial mean rotation of the surrounding flow [21]. It is defined as twice the intrinsic material rotation angle [22] and can be conveniently expressed as the trajectory integral of the instantaneous deviation of a fluid particle's vorticity from the flow's spatial mean vorticity [21]:

$$\mathrm{LAVD}_{t_0}^{t}(\boldsymbol{x}_0) = \int_{t_0}^{t} |\boldsymbol{\omega}(\boldsymbol{x}(\tau;\boldsymbol{x}_0),\tau) - \overline{\boldsymbol{\omega}}(\tau)| \mathrm{d}\tau \tag{5}$$

where $\boldsymbol{\omega}(\boldsymbol{x}(\tau;\boldsymbol{x}_0),\tau)$ is the local instantaneous vorticity at time $\tau$ of a fluid particle with initial position $\boldsymbol{x}_0$ advected from time $t_0$ to $t$, and $\overline{\boldsymbol{\omega}}(\tau)$ is the spatial mean vorticity at time $\tau$. In practice, the local vorticity is computed from velocity gradients, and the absolute deviation from the spatial mean is integrated along particle trajectories to obtain the LAVD. The vortex cores are identified as the peaks of the LAVD with which their locations can be tracked over time. In order to improve the calculation of the LAVD and the vortex core trajectories, the velocity fields and the vortex core trajectories were smoothed using a robust spline smoothing algorithm based on the discrete cosine transform and a penalized least squares approach [23]–[25]. LAVD was selected over traditional Eulerian criteria because it offers several crucial advantages for accurately identifying and defining fluid structures. These advantages include guaranteed material coherence, dynamic consistency, and computational robustness. In contrast, traditional Eulerian criteria tend to be non-objective, meaning their results can vary depending on the observer's reference frame. However, LAVD is objective, providing a consistent assessment of vortices regardless of coordinate rotations or translations. Peak vorticity, particularly measured through the local maxima of the LAVD, is significant as it identifies the singular center of a coherent vortex, which is explicitly defined as the vortex center. Additionally, in traditional Eulerian methods, defining a vortex boundary often depends on arbitrary thresholding and the selection of a reference frame [21], [22]. Some Reynolds numbers examined in this study fall within the range discussed by Palacios-Morales and Zenit [26], specifically 150 < Re < 260. For such a range, their study determined the limit stroke ratio to be between 4 and 6. In this current study, based on the analysis of the LAVD for the identification of the center, size and behavior of the vortices in the flow field following [21], the cases where a single vortex is generated are $\mathrm{t}_\mathrm{f}^*$ = 1, 2, 3, and 4, whereas $\mathrm{t}_\mathrm{f}^*$ = 5 includes trailing jets.

## III. RESULTS AND DISCUSSION

Three primary flow behaviors are identified in this study and are elucidated in Fig. 3. 1) In Fig. 3(a), we observe a **short-time decay** which occurs when the vortex rings propagate through the cavity (about midway for $t_f^*$ = 1, $s/d$ = 1.5) and dissipate before reaching the lower wall. 2) In Fig. 3(b), we observe a **decay at the lower wall**, which involves vortex rings traveling down to the cavity apex where they subsequently decay. 3) In Fig. 3(c), we observe a **wall rebound** that is characterized by the inner cores of the vortex rings reaching the cavity apex, undergoing stretching, and rebounding upward. A distinct subcategory of this wall rebound behavior features the formation of secondary vortices, as in Fig. 3(d), where the inner cores interact with the wall, rebound upward, and subsequently break up into smaller vortex structures. The light blue vortex-ring originates from the inner left ring, while the green and magenta vortex-rings come from the inner right vortex. Each of these flow behaviors is examined in greater detail in what follows. We nondimensionalize all spatial coordinates with respect to the jet diameter ($x^* = x/d$, $y^* = y/d$), velocities with respect to the mean jet velocity ($u^* = u/\overline{U}$, $v^* = v/\overline{U}$), and time as $t^* = t\overline{U}/d$.

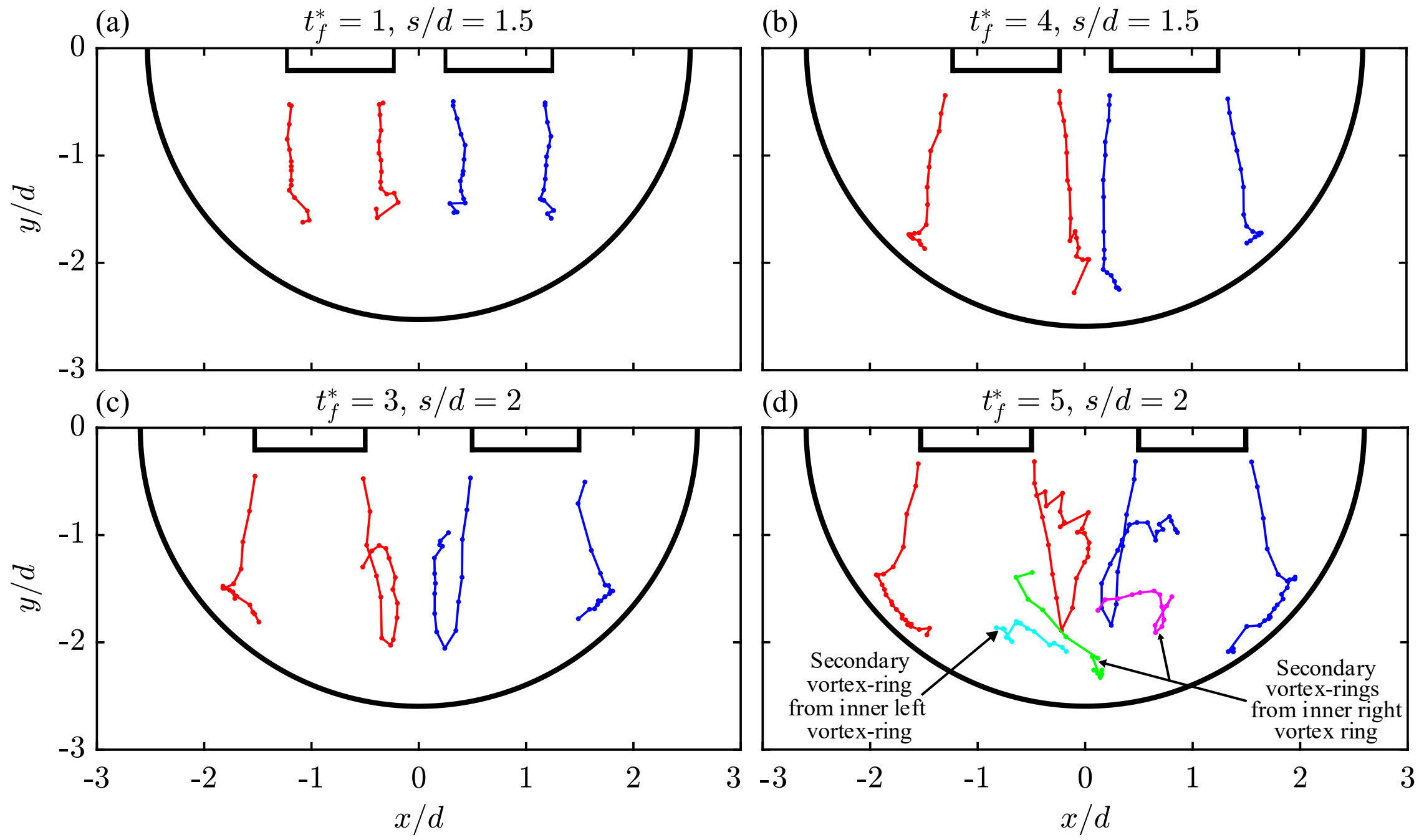

**FIG 3.** Temporal evolution of the position of the vortex cores for the three identified primary flow behaviors: (*a*) short time decay, (*b*) decay at the lower wall, (*c*) wall rebound, and wall rebound with secondary vortex generation, where the light blue vortex-ring originates from the inner left vortex ring (red) and the green and magenta originate from the inner right vortex-ring (blue) (d – see main text for more details).$t_f^*$: formation time; $s/d$: spacing ratio between the vortex rings.

## A. Short-time decay

The results for $t_f^*$ = 1 and $s/d$ = 1.5 are used to illustrate the behavior of the short-time decay regime, though we observe this behavior for all investigated spacings for the same formation time. Fig. 4(a) and 4(b), respectively, show the LAVD at the start of the formation of the vortex rings ($t^*$ = 0.46) and when the vortex rings reach their lowest point ($t^*$ = 1.96). The twin vortices emerging into the hemisphere are advected towards the apex of the cavity but dissipate before ever reaching the lower wall. In Fig. 4(c), we show the temporal evolution of axial (vertical) velocity $v^*$ taken at $y/d$ = -1.38, which is just downstream of the mid-radius of the hemisphere ($y/d$ = -1.27). The axial velocity is characterized by rather small magnitudes throughout the entire measurement duration, given the less energetic flow and the decay within a shorter time. At around $t^*$ = 1.5, the vortex rings reach $y/d$ = -1.38 and remain there as they decay, which is marked by the four persistent patterns of axial velocity sign changes in Fig. 4(c) and the significant fading of axial velocity magnitude within about two formation times ($2t_f^*$). The normalized vorticity magnitude ($\omega^* = \omega d/\overline{U}$) of the four identified cores, using the peaks of the LAVD, is plotted over time in Fig. 4(d). The peak core vorticity $|\omega^*_{\max}|$ is about 7.8 for all cores and occurs at around $t^*$ = 0.82. It is interesting to note that the peak vorticity does not occur at $t^* = t_f^*$. This is due to the interaction between the two vortex rings during the ejection period, which eventually limits the entrainment of vorticity from the shear layer. We subsequently note an exponential decay (i.e., of the form $e^{-\beta t^*}$) with decay rate $\beta$ = 0.71 in the vorticity starting at $t^*$ = 0.82 for $t_f^*$ = 1 and $s/d$ = 1.5. The vorticity falls to about 25% of its peak after about two formation times (i.e., by $t^*$ = 2.45). Although the two inner parts of the vortex rings slightly lag behind the outer parts due to their interaction as they are advected downstream (see Fig. 4(b)), the evolution of vorticity for all four cores in the plane of measurement is nearly identical, demonstrating that the vortex rings maintain their symmetry as they propagate downstream. The same symmetry is observed for the other cases falling in the short-time decay regime, namely for $t_f^* = 1$ and $s/d = 2.0$, 2.5, and 3.0.

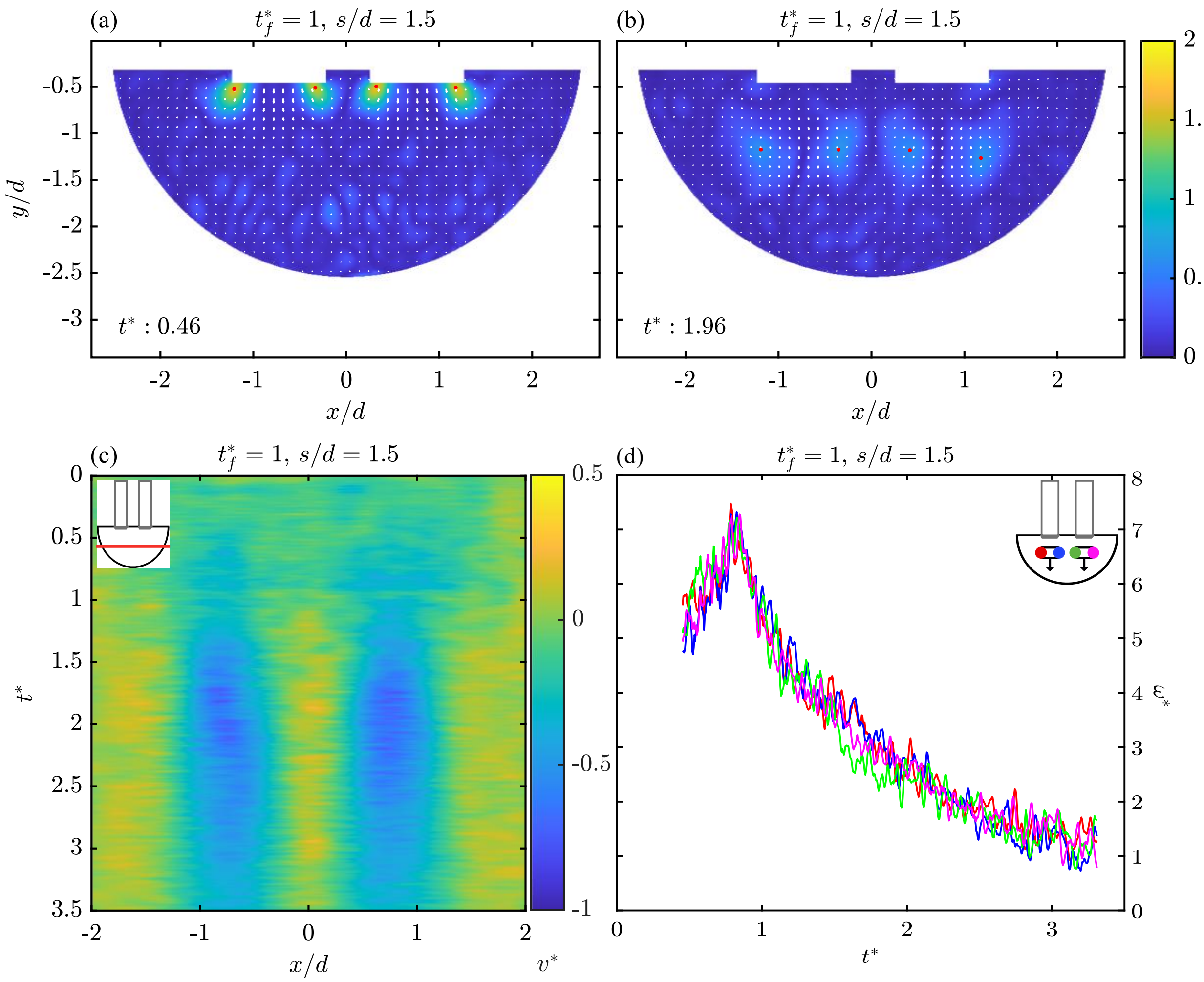

**FIG 4.** Characteristics of the short-time decay behavior illustrated using the case $t_f^* = 1$ and $s/d = 1.5$. Lagrangian-averaged vorticity deviation at (*a*) $t^* = 0.46$ and (*b*) $t^* = 1.96$. (*c*) Temporal evolution of the axial velocity profile at $y/d = -1.38$, near the hemisphere center. (*d*) Temporal evolution of the absolute core vorticity ($|\omega^*|$). $t_f^*$: formation time; $s/d$: spacing ratio between the vortex rings.

## B. Decay at the lower wall

In this regime, the inner cores of the vortex rings advect down to the lower wall and decay there. In particular, we observe this behavior under three different scenarios. The first scenario occurs when the formation time, an indicator of jet strength, permits just enough energy to be imparted to the vortex rings such that they just reach the lower wall. We observe this behavior at a formation time of $t_f^* = 2$ for all investigated spacing ratios ($s/d \geq 1.5$). This scenario can also be considered as the limiting case of the short-time decay behavior. The second scenario occurs when the formation time permits more than enough energy to be imparted to the vortex rings to have them reach the lower wall; however, the spacing ratio is sufficiently small to result in a strong interaction between their inner cores that dissipates enough energy to limit any dynamic interaction with the lower wall once they reach it. By dynamic interaction, we are referring to the dynamics observed in the third ‘wall rebound’ regime that we will discuss later. For higher formation times in this scenario, symmetry-breaking interactions between the two inner vortex cores can nonetheless occur at the lower wall. In our study, this scenario occurs for formation times of $t_f^* > 2$ only for the lowest investigated spacing ratio ($s/d$ = 1.5). Lastly, the third scenario occurs when the formation time again permits the vortex rings to reach the lower wall, however the spacing ratio is sufficiently large such that the inner cores first reach the lateral wall and subsequently advect toward the apex following the curvature of the hemisphere, dissipating enough energy through viscous effects with the wall to limit any further dynamic interactions once they reach the apex. We observe this scenario for formation times of $2 < t_f^* < 5$ at a spacing ratio of $s/d$ = 3.0.

The fluid dynamics of the first and third scenarios described above are less interesting, again with the first representing a limiting case of the short-time decay behavior. Therefore, in what follows, we use the results for $t_f^*$ = 4 and $s/d$ = 1.5 to represent the second scenario described above, namely, for decay at the lower wall behavior with inner core interactions at close spacing ratios. This particular case also demonstrates symmetry-breaking inner vortex core interaction at the lower wall. Fig. 5(a) and 5(b) respectively show the two vortex rings (using LAVD magnitude) as they approach the lower wall and the beginning of their subsequent interaction. The twin vortex rings maintain their symmetry with respect to the $x/d$ = 0 plane as they cross the mid-radius plane of the hemisphere at approximately $t^*$ = 3.25. The outer cores reach the lateral wall shortly after ($t^*$ = 4.0), while the inner cores continue to extend down to the apex of the cavity. At the larger spacing ratio ($t_f^*$ = 4, $s/d$ = 3.0) exhibiting the third scenario, the outer cores naturally reach the lateral wall earlier ($t^*$ = 2.4) while the inner cores reach the apex much later during their decay (at around $t^*$ = 5.0).

In all investigated cases, upon reaching the lateral wall, the outer cores remain in a relatively fixed position and ultimately dissipate as a result of their interaction with the wall. Meanwhile, the inner cores may interact at the apex of the hemisphere symmetrically ($t_f^*$ = 3, $s/d$ = 1.5) or asymmetrically ($t_f^*$ = 4, $s/d$ = 1.5). In Fig. 5(b), we observe a symmetry-breaking interaction between the inner cores as one core outcompetes the other. At the lowest spacing in this regime, this behavior was only observed for a formation time of $t_f^*$ = 4 in all 3 realizations of the same experiment, indicating that a shear layer instability may be manifesting between the two inner cores. Fig. 5(c) shows the temporal evolution of axial velocity taken at $y/d$ = -1.38. In contrast to the short-time decay behavior in Fig. 4(c), the vortex rings continue to advect past $y/d$ = -1.38, and therefore, the velocity decay cannot be directly observed in Fig. 5(c). We note however a rather narrow portion of upward flow between $t^*$ = 2.0 and 5.0, demonstrating that the two inner cores press against each other and stretch as they are advected downstream (see also Fig. 5(a)), which was not apparent at the same spacing ratio for $t_f^*$ = 1. We observe the same vortex stretching process at the same spacing ratio ($s/d$ = 1.5) for formation times of $t_f^* \geq 2$. For the larger spacing ratio ($s/d$ = 1.5) exhibiting inner core decay at the lower wall, the inner cores are sufficiently far away to have limited interaction between each other (no mutual vortex stretching observed).

A distinctive feature in the second scenario of the lower wall decay regime is the strong initial interaction between the two inner cores and possible symmetry-breaking interactions at the apex. Fig. 5(d) illustrates this asymmetry through the temporal evolution of the absolute core vorticity in the plane of measurement ($|\omega^*|$). The early stage of vortex ring development remains rather symmetric, though the inner cores attain a larger first vorticity peak of 8.2 at $t^*$ = 1.6, compared to 7.0 for the outer cores. This is followed by a decay of vorticity from $t^*$ = 6.0 to $t^*$ = 14.2 until the inner cores reach the apex. The two inner cores interact, breaking the symmetry as one core entrains vorticity from the other. The dynamics of this process is shown through the insets in Fig. 5(d). The inner core on the right subsequently exhibits a sharp increase in core vorticity as a result, peaking at 13.5 at around $t^*$ = 5.7, whereas the inner core on the left peaks at 10.6 slightly earlier around $t^*$ = 5.5 before it begins to be sheared out. The other cores exhibit a lesser increase in vorticity after the formation time up to $t^*$ = 5.0. Interestingly, all vortex cores again decay significantly within about two formation times ($2t_f^*$) from their peak, with the high vorticity inner core therefore decaying within a shorter time than the others. However, the exponential decay rates of all cores are similar, with that of the inner cores being $\beta$ = 0.33 and the outer cores $\beta$ = 0.32.

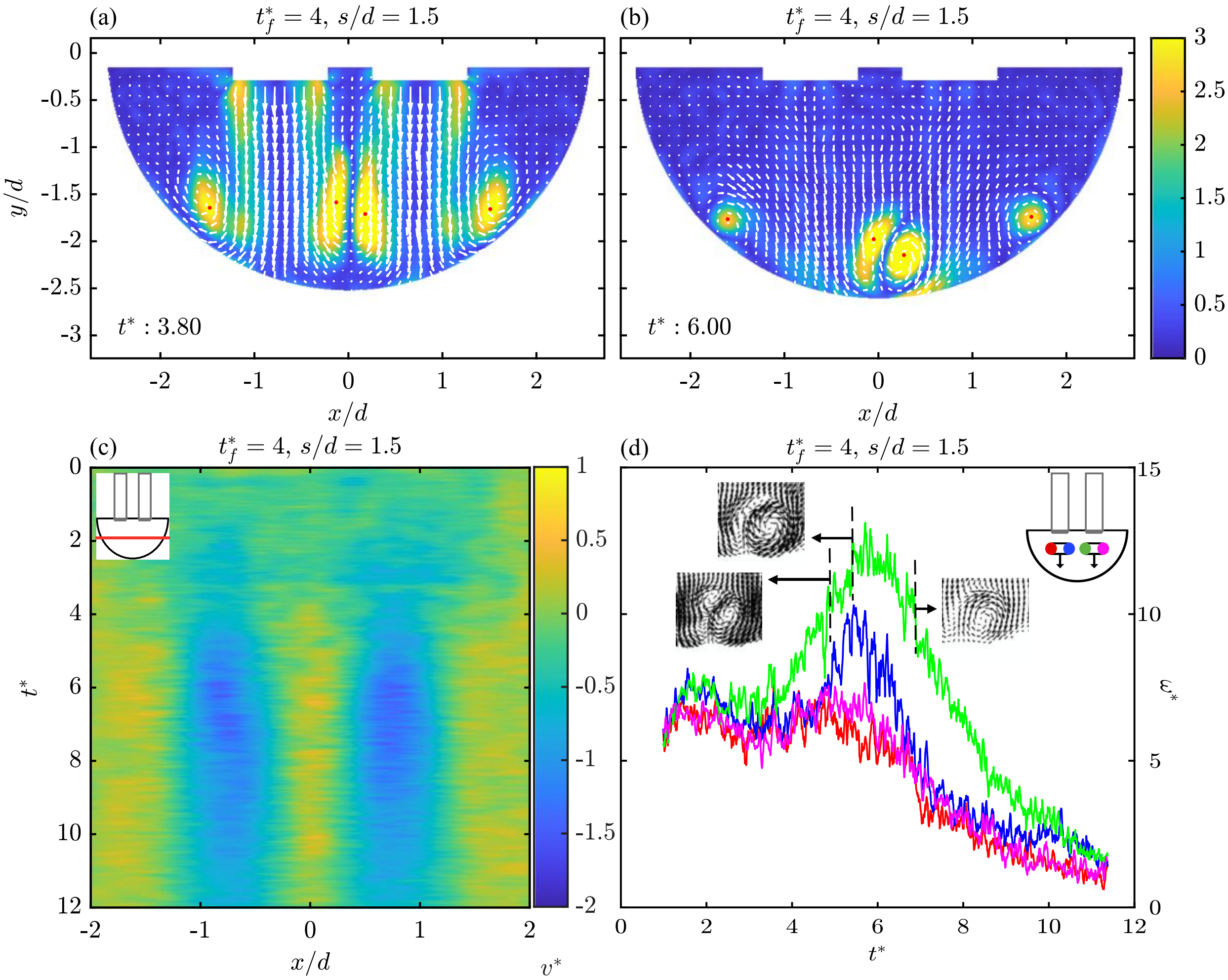

**FIG 5.** Characteristics of the decay at the lower wall behavior illustrated using the case $t_f^*$ = 4 and $s/d$ = 1.5. Lagrangian-averaged vorticity deviation at (*a*) $t^*$ = 3.25 and (*b*) $t^*$ = 6.00. (*c*) Temporal evolution of the axial velocity profile at $y/d$ = -1.38, near the hemisphere center. (*d*) Temporal evolution of the absolute core vorticity ($|\omega^*|$). The insets demonstrate the interaction and transfer of vorticity between the two inner cores at the apex of the hemisphere. $t_f^*$: formation time; $s/d$: spacing ratio between the vortex rings.

### C. Wall rebound

In this regime, the inner cores of the vortex rings not only reach the lower wall, but also interact with the wall and exhibit a deflection upward within the cavity. Two scenarios are observed in this regime, namely, a symmetric upward propagation of the inner cores and an asymmetric upward propagation marked by the generation of secondary vortices. The case corresponding to $t_f^*$ = 3 and $s/d$ = 2 is used to represent the wall rebound behavior with a symmetric upward propagation of the inner cores. The results are displayed in Fig. 6.

The inner vortex cores reach the bottom of the hemisphere at around $t^*$ = 4.07, as seen from the LAVD magnitude in Fig. 6(a). They then stretch vertically and rebound up toward the center of the cavity as they decay (Fig. 6(b)). In some cases where the upward propagation is symmetric, the inner cores rebound up to the mid-radius of the cavity (e.g., for $t_f^*$ = 3). In others, the inner cores can even reach the upper portion of the hemisphere (the nozzles). The outer cores decay at the lateral wall, as observed for the decay at the lower wall behavior. In Fig. 6(c), showing the evolution of axial velocity at $y/d$ = -1.38, the initial evolution is similar to the decay at the lower wall behavior in Fig. 5(c). The wall rebound behavior of the inner cores is then observed through a persistent positive upward velocity component at $x/d$ = 0 for larger times. Moreover, the upward velocity is seen to spread outward from $x/d$ = 0, indicating that the two cores begin to repel each other during their upward trajectory (see also Fig. 3(c) and 3(d)). From the temporal evolution of the absolute normalized vorticity (Fig. 6(d)), the outer vortex cores are observed to start decaying at the end of ejection ($t^*$ = $t_f^*$) due to their interaction with the lateral wall with an exponential decay rate of $\beta$ = 0.24 in Fig. 6(d). The inner cores, on the other hand, grow in vorticity as they rebound toward the center of the hemisphere, attaining a peak vorticity of 7.5 at $t^*$ = 5.45, and subsequently decaying at around $t^*$ = 7.67 with a decay rate of $\beta$ = 0.18.

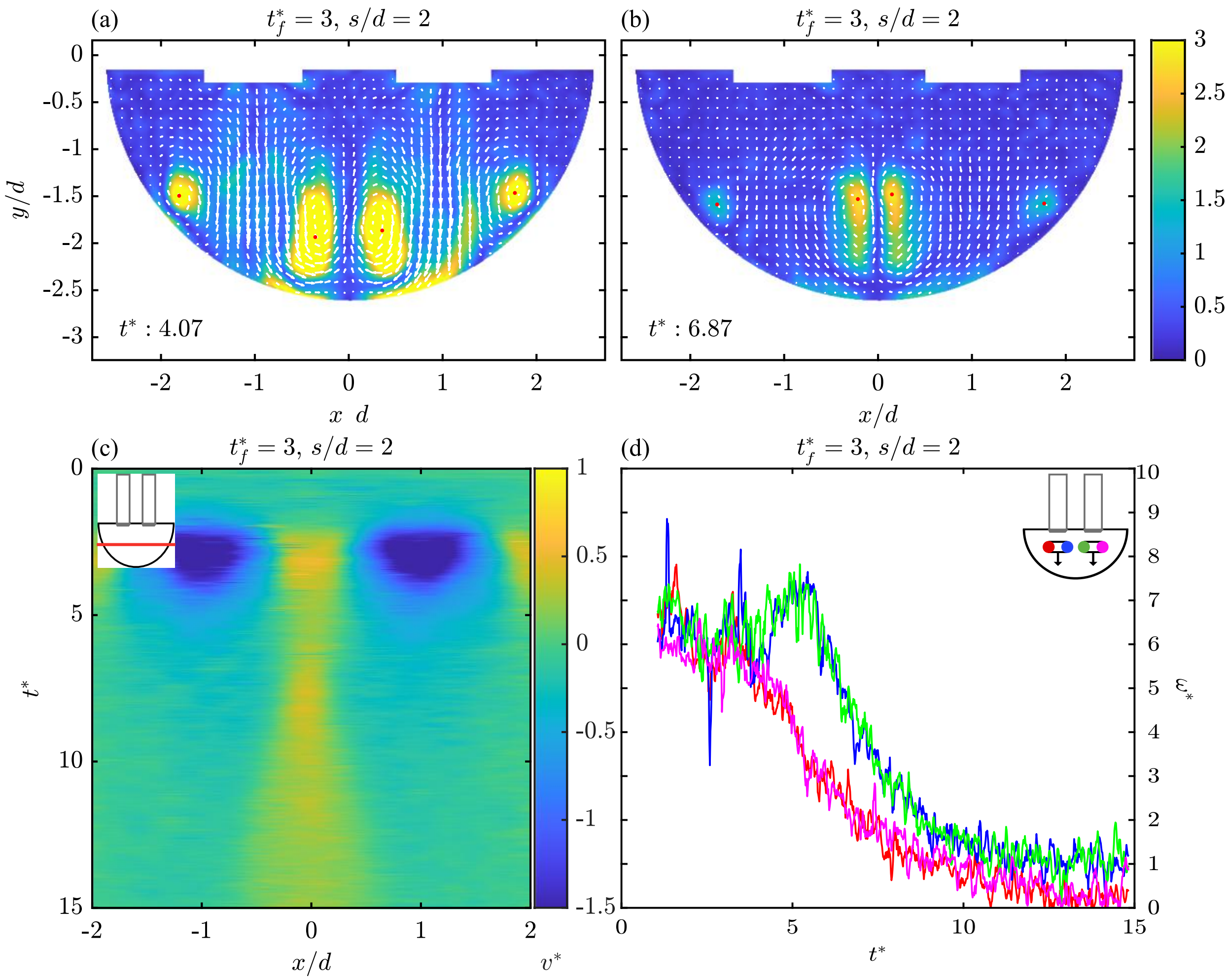

**FIG 6.** Characteristics of the wall rebound behavior illustrated using the case $t_f^*$ = 3 and $s/d$ = 2.0. Lagrangian-averaged vorticity deviation at (*a*) $t^*$ = 4.07 and (*b*) $t^*$ = 6.87. (*c*) Temporal evolution of the axial velocity profile at $y/d$ = -1.38, near the hemisphere center. (*d*) Temporal evolution of the absolute core vorticity ($\omega^*$). $t_f^*$: formation time; $s/d$: spacing ratio between the vortex rings.

For specific combinations of $t_f^*$ and $s/d$, such as $t_f^*$ = 5 with $s/d$ = 2.0, 2.5, and 3.0, the wall rebound behavior displays an additional feature characterized by the appearance of secondary vortices. In Fig. 7, the LAVD magnitude is shown for the spacing ratio of $s/d$ = 2.0 at the time when the inner cores reach the apex ($t^*$ = 5.05, Fig. 7(a)) and at a time where their upward trajectory is accompanied by the generation of the secondary vortex cores ($t^*$ = 7.80, Fig. 7(b)). Within the plane of measurement, the secondary vortex generation process has elements that would indicate a shear layer instability. Specifically, for these two combinations, the two inner cores are close together when they reach the apex and rebound upward, creating a strong and narrow layer of upward velocity between them and, consequently, a strong local unstable shear layer that can ultimately break the symmetry of the flow. Though we did not capture the three-dimensional flow, the secondary vortices seem to occur as a vortex ring splitting from the inner portions of the two main vortex rings.

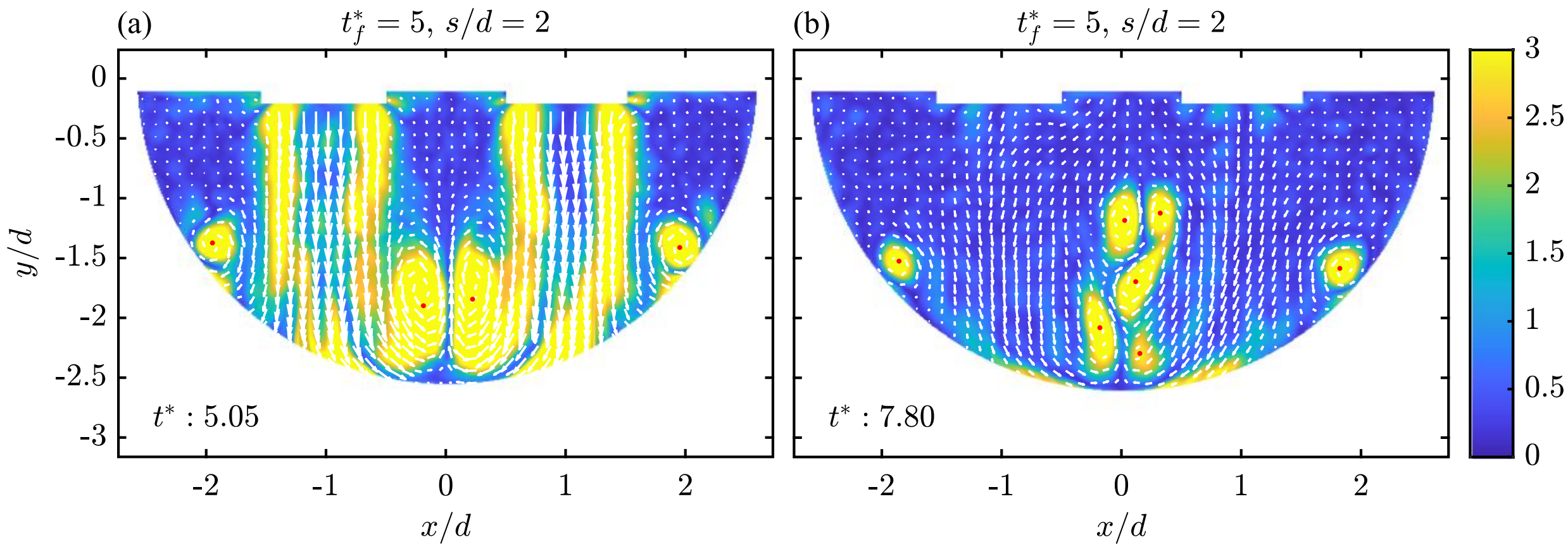

**FIG 7.** Lagrangian-averaged vorticity deviation for the case $t_f^*$ = 5 and $s/d$ = 2.0 at (*a*) $t^*$ = 5.05 and (*b*) $t^*$ = 7.80. In addition to wall rebound behavior, the emergence of secondary vortices is observed.

### D. Flow behavior map

Fig. 8 shows the vertical displacement of the inner right vortex cores for all cases beginning from $t^*$ = 1.0. For those cases exhibiting the short-time decay behavior (Fig. 8(a)), the inner vortex cores do not reach the lower wall of the cavity ($y/d$ = -2.2). This behavior is only observed at the lowest value of formation time and appears to be independent of $s/d$. The inner vortex trajectories for all cases follow a linear path with a slope of -0.35, representing the induced velocity of the inner cores ($v_v^*$). As for the outer cores, they either reach the lateral wall and decay there (at higher $s/d$ values) or decay near the center of the cavity (at lower $s/d$ values). The vertical displacement of the inner vortex cores for all cases exhibiting the decay at the lower wall behavior is shown in Fig. 8(b). All cases exhibit a downward displacement with the same induced core velocity magnitude of $v_v^*$ = 0.43, despite the different formation times between cases. This result is interesting as it indicates that a limit on the induced inner core velocity is attained at rather low formation times (as low as $t_f^*$ = 2). The effects of the confined elastic environment and the vortex interactions thus result in significantly different behavior from what is observed for the single pulsed jet in an open environment where the propagation speed can reach higher values for lower Reynolds numbers [26]. Fig. 8(c) shows the inner vortex core trajectories for all cases exhibiting the wall rebound behavior. Again, all cases show an initial downstream with a similar core propagation speed of 0.41. Except for the highest formation time ($t_f^*$ = 5), the inner cores reach their lowest point at roughly the same time ($t^*$ = 4.8). The higher formation times ($t_f^*$ = 4 and 5) are characterized by more expansion of the elastic hemisphere (i.e., larger injection volumes), causing the apex to be displaced downward. It can clearly be observed that after reaching the lower wall of the hemisphere (the apex being at $y/d$ = -2.2), the inner vortex cores rebound upward and reach the upper portion of the hemisphere near the nozzles for some combinations of formation time and spacing ratio such as ($t_f^*$, $s/d$) = (4, 2.5), (5, 2.5), and (5, 3.0). The type of interaction between the jet and the silicone membrane, i.e., rigid/moving wall, depends mainly on $t_f^*$. For low $t_f^*$ values (1 to 3), the inner cores of the twin jets either never reach the wall (short-time decay behavior) or reach the wall after the silicone membrane has stopped expanding (i.e., interaction with a rigid wall). For higher $t_f^*$ values, the inner cores reach the wall while the membrane is still expanding (i.e., interaction with a moving wall).

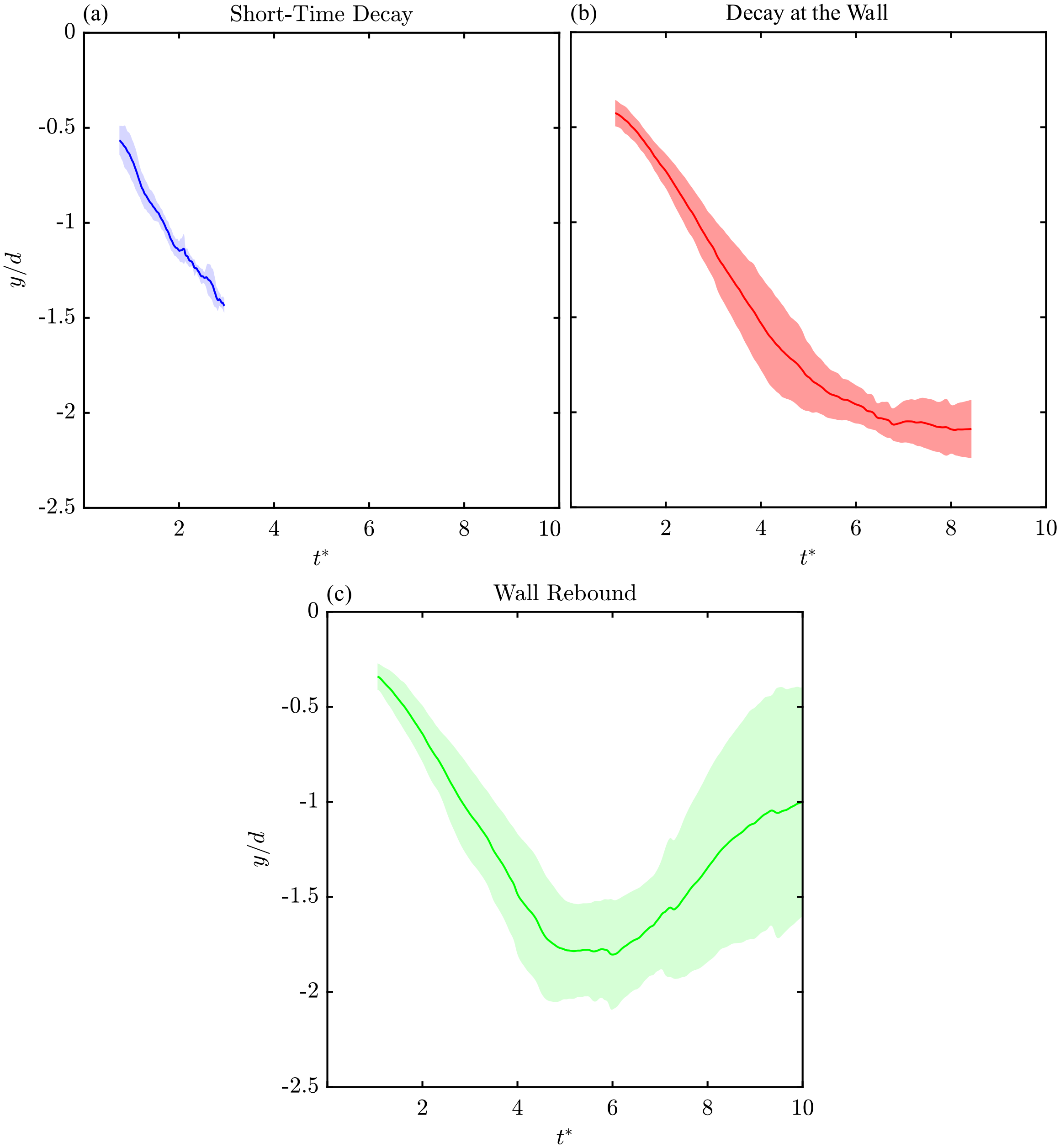

**FIG 8.** Temporal evolution of the axial position ($y/d$) of the inner core of the vortex rings for all cases exhibiting (*a*) short-time decay behavior, (*b*) decay at the lower wall behavior, and (*c*) wall rebound behavior. The trajectories of the vortex cores are tracked by following the location of the peak of the LAVD. $t_f^*$: formation time; $s/d$: spacing ratio between the vortex rings. The solid line represents the mean value of the core trajectories while the shaded regions indicate the lower and upper bounds of the core trajectory standard deviation.

Fig. 9 summarizes the results for all the cases investigated in this study by showing the different behaviors as a function of the formation time ($t_f^*$) and the spacing ratio ($s/d$). The short-time decay behavior (Fig. 9, blue) is observed only at the lowest formation time ($t_f^* = 1$), independently of the spacing ratio ($s/d$). In these cases, the vortices possess insufficient energy to reach the lower wall of the hemisphere and instead dissipate near the cavity centerline. At higher formation times, the inner parts of the vortex rings have enough energy to reach the lower wall of the hemisphere. The inner cores then either decay at the lower wall (Fig. 9, red) or rebound upward (Fig. 9, green). At a formation time of $t_f^* = 2$, the vortex rings have just enough energy to reach the lower wall and a rebound is never observed. For particularly low spacing ratios ($s/d = 1.5$), again, no wall rebound is observed. The inner vortex cores interact very strongly as they emerge and propagate downstream, dissipating significant energy as they reach the lower wall and ultimately decay there. For sufficiently high formation times with spacing ratios that limit the initial interaction of the vortex rings, the vorticity of the inner cores as they reach the lower wall is sufficient to cause the inner parts of the vortex rings to stretch vertically and induce an upward motion of the inner cores.

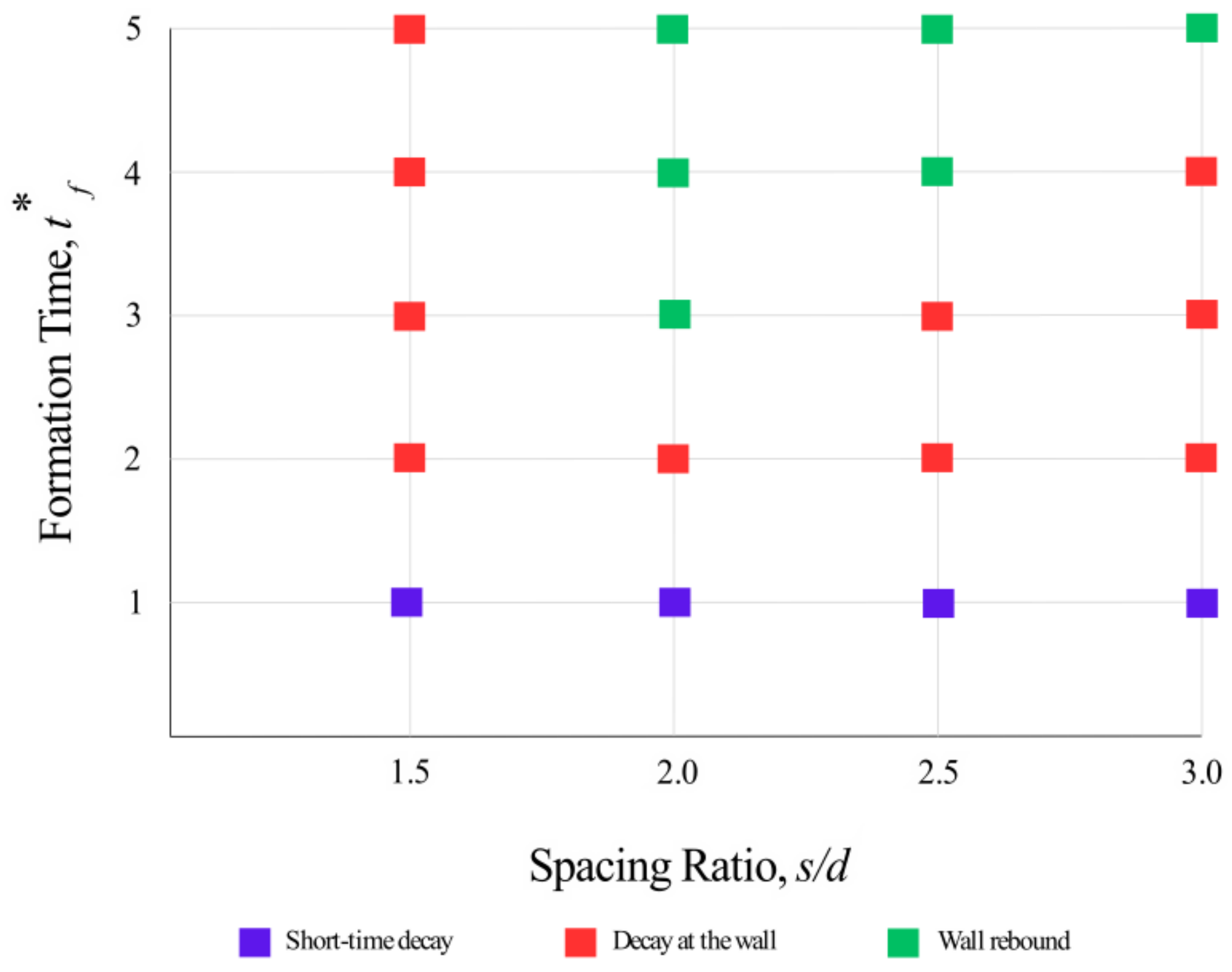

**FIG 9.** Flow map illustrating the different flow configurations as a function of $t_f^*$ and $s/d$. Short-time decay is represented by blue squares, Decay at the wall is represented by red squares and Wall rebound is represented by green squares.

## IV. CONCLUSION

Most existing studies on vortex-wall interactions have focused on vortex-ground interactions relevant to aircraft takeoff and landing. The interaction of a single vortex ring with a wall has been extensively investigated under various configurations, including flat plates [27], static and rotating cylindrical surfaces [28], V-walls [10], and convex/concave boundaries [10], [29]. These studies have significantly advanced our understanding of collision mechanisms, counter-vorticity generation during wall approach, and vortex breakup dynamics. While building on these foundations, the present study addresses a different and fundamentally more challenging problem inspired by pathological left ventricular flow, specifically, aortic or pulmonary valve regurgitations during diastole, where twin vortex rings emerge simultaneously into the deforming heart cavities. Here, we investigate the interaction of twin vortex rings generated by the expansion of a hemispherical cavity, introducing unique hydrodynamic considerations absent in prior single-vortex or simplified geometric studies. In particular, our results highlight the emergence of three distinct behaviors: short-time decay, decay at the lower wall, and wall rebound.

For the lowest examined formation time of $t_f^*$ = 1, only the short-time decay behavior is observed (Fig. 9, blue). The two vortex rings remain symmetric about $x/d$ = 0 as they propagate downstream, though they never reach the lower wall of the hemisphere. The vortex rings interact only weakly, causing a slight lag in the downstream displacement of the inner cores relative to the outer cores. The vorticity of all cores nonetheless exhibits the same behavior, peaking slightly before the formation time and subsequently decaying at the same exponential decay rate, with the vorticity falling to within 25% of its peak after about two formation times. In this regime, the inner cores of the vortex rings have a rather constant induced propagation speed that can still increase should the formation time increase.

The lower wall decay behavior encompasses a wider range of conditions. This behavior appears when the vortex rings have just enough energy to reach the lower wall ($t_f^*$ = 2 for all $s/d$ values) or when they lose part of their energy through mutual interaction at close spacings ($s/d$ = 1.5). At the lowest spacing ratio ($s/d$ = 1.5) and for higher formation times, the inner cores still reach the lower wall in close proximity; however, with sufficient vorticity to cause them to interact strongly, sometimes breaking the symmetry of the flow about $x/d$ = 0 as one core outcompetes the other. Interestingly, this behavior persists at larger spacings ($s/d$ = 3), but only for intermediate formation times ($2 \leq t_f^* \leq 4$). In these cases, the inner cores of the vortex rings reach the lower wall through a different trajectory. Given the larger spacing ratio, they first reach the lateral wall and subsequently follow the curvature of the hemisphere toward the apex. In this regime, the induced propagation speed of the inner cores appears to have attained a physical limit which is maintained in the wall rebound regime.

The wall rebound behavior emerges under specific conditions that balance vortex strength and spacing. In this work, it is observed for intermediate spacings ($s/d$ = 2 and 2.5) with sufficiently high formation times ($t_f^*$ > 3). Similarly, strongly energetic vortex rings ($t_f^*$ = 5) at large spacings ($s/d$ = 3) also demonstrate the wall rebound behavior. In these cases, the vortex rings maintain enough separation to limit mutual interaction while retaining sufficient energy to interact with the hemispherical cavity walls, rebounding in a manner analogous to vortex collisions with rigid flat or curved surfaces.

This fundamental study provides a deeper understanding of the flow dynamics of twin pulsed jets in an elastic cavity. The findings have potential applications in improving our current understanding of the complex flow patterns generated by certain medical devices and pathological conditions, such as heart valve regurgitation and heart valve repair, and their impact on cardiac function and performance. Future research should investigate the three-dimensional aspects of the flow and explore a wider range of jet configurations, including non-parallel and asynchronous twin pulsed jets.

## ACKNOWLEDGMENTS

HDN acknowledges the support of the Natural Science and Engineering Research Council of Canada/Conseil de la Recherche en Sciences Naturelles et en Génie du Canada (RGPIN-2023-03916). LK acknowledges the support of the Natural Science and Engineering Research Council of Canada/Conseil de la Recherche en Sciences Naturelles et en Génie du Canada (RGPIN-344164-07).
GDL, HDN and LK conceived the study. LSM designed the experiment. LSM, WS and GDL performed the measurements. LSM and GDL post-processed the data. LSM, HDN and LK wrote the manuscript. All authors participated in revising the manuscript and read and approved the final manuscript.

## DATA AVAILABILITY

The data that support the findings of this article are openly available [30].